\documentclass[a4paper,11pt]{article}
\usepackage[left=3.17cm,right=3.17cm,top=2.54cm,
headheight=0.5cm,headsep=0.54cm,bottom=2.54cm,footskip=0.79cm
]{geometry}

\usepackage[all]{xy}
\usepackage{amsmath,amsfonts,amssymb,amsthm,epsfig,amscd,comment,latexsym,psfrag}
\usepackage{nicefrac,xspace,tikz}

\usetikzlibrary{arrows}
\usetikzlibrary{decorations.markings}
\def\Z{\mathbb Z}

\def\C{\mathbb C}

\def\1{{\bf{1}}}

\usepackage[pdfstartview=FitH]{hyperref}

\def\3Dtwoboxy{\begin{tikzpicture}
\draw (0,0)rectangle(0.5,0.25);
\draw (0.25,0) -- (0.25,0.25) [-];
\draw (0,0.25) -- (0.175,0.35) [-];
\draw [shift = {+(0.25,0)}](0,0.25) -- (0.175,0.35) [-];
\draw [shift = {+(0.5,0)}](0,0.25) -- (0.175,0.35) [-];
\draw [shift = {+(0.5,-0.250)}](0,0.25) -- (0.175,0.35) [-];
\draw (0.175,0.35) -- (0.675,0.35) [-];
\draw [shift = {+(0.25,0)}] (0.425,0.35) -- (0.425,0.1) [-];
\end{tikzpicture}}

\def\footnoterule{\kern 1mm \hrule width 7cm \kern 2.2mm}%

\newcommand{\bea}{\begin{eqnarray}}
\newcommand{\eea}{\end{eqnarray}}
\newcommand{\beaa}{\begin{eqnarray*}}
\newcommand{\eeaa}{\end{eqnarray*}}
\newcommand{\be}{\begin{equation}}
\newcommand{\ee}{\end{equation}}
\newcommand{\nn}{\nonumber}

\begin{document}

\title{Jack polynomials, $\hbar$-dependent KP hierarchy and affine Yangian of ${\mathfrak{gl}}(1)$}
\author{Wang Na\dag\footnote{Corresponding author: wangna@henu.edu.cn },\ Zhang Can\dag\ Wu Ke\ddag\\
\dag\small School of Mathematics and Statistics, Henan University, Kaifeng, 475001, China\\
\ddag\small School of Mathematical Sciences, Capital Normal University, Beijing 100048, China}

\date{}
\maketitle

\begin{abstract}
In this paper, we discuss the relations between the Jack polynomials, $\hbar$-dependent KP hierarchy and affine Yangian of ${\mathfrak{gl}}(1)$. We find that $\alpha=\hbar^2$ and $h_1=\hbar, \ h_2=-\hbar^{-1}$, where $\alpha$ is the parameter in Jack polynomials, and $h_1,\ h_2$ are the parameters in affine Yangian of ${\mathfrak{gl}}(1)$. Then the vertex operators which are in Jack polynomials are the same with that in $\hbar$-KP hierarchy, and the Jack polynomials can be used to describe the tau functions of the $\hbar$-KP hierarchy.
\end{abstract}
\noindent
{\bf Keywords: }{$\hbar$-KP hierarchy, Affine Yangian, Jack polynomials, vertex operators, Boson-Fermion correspondence.}

\section{Introduction}\label{sect1}
The KP hierarchy is one of the most important integrable hierarchies and it arises in many different fields of mathematics and physics such as enumerative algebraic geometry, topological field and string theory\cite{DKJM, MJD}. Meanwhile Young diagrams and symmetric functions are of interest to
many researchers and have many applications in mathematics including combinatorics
and representation theory of the symmetric and general linear group\cite{FH,Mac,weyl,Stan}. Schur functions can be used to describe the tau functions of the KP hierarchy, and the vertex operators which realize the Schur functions have close relations with the Fermions in the KP hierarchy. In this paper, we generalize these to the case of Jack polynomials and $\hbar$-KP hierarchy.

In \cite{TT1,TT2}, the authors K. Takasaki and T. Takebe defined the $\hbar$-dependent KP hierarchy ($\hbar$-KP hierarchy for short) by introduced a formal parameter $\hbar$. It is a generalization of the KP hierarchy in the sense that it becomes the KP hierarchy when $\hbar\rightarrow 1$. When $\hbar\rightarrow 0$, the $\hbar$-KP hierarchy becomes the dispersionless KP hierarchy. The $\hbar$-KP hierarchy was introduced to study the dispersionless KP hierarchy \cite{TT1}. The $\hbar$-KP hierarchy is defined by the Lax representation
\[
\hbar \frac{\partial L}{\partial x_j}=[B_j,L],\ \ \text{with}\ \ B_j=(L^j)_+,
\]
where the Lax operator $L$ is the pseudodifferential operator of the following form
\[
L=\hbar\partial+\sum_{j=1}^\infty f_j (\hbar\partial)^{-j}.
\]

In \cite{deformedKPJack}, we introduce the vertex operators
\beaa
X_+(z)&=&\sum_{n\in\Z}X_n^+ z^n=\exp\left(\sum_{n\geq 1}\frac{x_n}{\sqrt{\alpha}}z^{n}\right)\exp\left(-\sum_{n\geq 1}\frac{\partial_{x_n}}{n}\sqrt{\alpha}z^{-n}\right)\\
X_-(z)&=&\sum_{n\in\Z}X_n^- z^n=\exp\left(-\sum_{n\geq 1}\frac{x_n}{\sqrt{\alpha}}z^{n}\right)\exp\left(\sum_{n\geq 1}\frac{\partial_{x_n}}{n}\sqrt{\alpha}z^{-n}\right).
\eeaa
They realize the Jack polynomials $\tilde{J}_\lambda$, and satisfy the Fermion relations. Then we define an integrable hierarchy by
the bilinear relations:
\be
\sum_{m+n=-1} X_{m}^-\tau\otimes X_n^+\tau=0.
\ee

In this paper, we show that the integrable hierarchy above is exactly the $\hbar$-KP hierarchy by the Hirota form. Then we give the Boson-Fermion correspondence of $\hbar$-KP hierarchy, and describe the tau functions of $\hbar$-KP hierarchy by using the Jack polynomials $\tilde{J}_\lambda$ and $S_\lambda\left(\frac{x}{\sqrt{\alpha}}\right)$.

 The paper is organized as follows. In section \ref{sect2}, we recall the definition of $
\hbar$-KP hierarchy. In section \ref{sect3}, we recall the definitions of the affine Yangian of ${\mathfrak{gl}}(1)$ and the Jack polynomials. Then we show the properties of the Jack polynomials and the vertex operators. In section \ref{sect4},  we use the Hirota equations to show that the $\hbar$-dependent KP hierarchy is exactly the integrable hierarchy defined in \cite{deformedKPJack}. In section \ref{sect5}, we give the Boson-Fermion correspondence in the $\hbar$-KP hierarchy.
\section{$\hbar$-dependent KP hierarchy}\label{sect2}
In \cite{TT1,TT2}, the authors K. Takasaki and T. Takebe defined the $\hbar$-dependent KP hierarchy ($\hbar$-KP hierarchy for short) by introduced a formal parameter $\hbar$. When $\hbar\rightarrow 0$, the $\hbar$-KP hierarchy becomes the dispersionless KP hierarchy, and when $\hbar\rightarrow 1$, it becomes the KP hierarchy. The tau functions and the wave function in $\hbar$-KP hierarchy are functions of parameters $x,t_1,t_2,\cdots$, while $x$ is only emerged in $x+t_1$. In this paper, we want the parameters are $x=(x_1,x_2,\cdots)$, which correspond to $(t_1,t_2,\cdots)$ in \cite{TT1,TT2}.

Consider a pseudodifferential operator
\be
L=\hbar\partial+\sum_{j=1}^\infty f_j (\hbar\partial)^{-j},
\ee
and the corresponding eigenvalue problem
\be\label{lwzw}
Lw=zw,
\ee
where $\partial=\frac{\partial}{\partial x}$.

We consider a formal solution
\bea
w&=&e^{\sum_{j=1}^\infty \frac{x_j}{\hbar}z^j}(1+\frac{w_1}{z}+\frac{w_2}{z^2}+\cdots)\\
&=& (1+{w_1}{(\hbar \partial)^{-1}}+{w_2}{(\hbar \partial)^{-2}}+\cdots)e^{\sum_{j=1}^\infty \frac{x_j}{\hbar}z^j}.\label{wM}
\eea
Then let
\be
M=1+\sum_{j=1}^\infty{w_j}{(\hbar \partial)^{-j}}.
\ee

Consider the linear system of equations
\be\label{hpartialwbjw}
\hbar\frac{\partial w}{\partial x_j}=B_jw,\ \ \text{with}\ \ B_j=(L^j)_+,
\ee
where $(L^j)_+$ is the differential operator part of $L^j$, that is, $(L^j)_+$ includes the terms $\partial^k, \ k\geq 0$ in $L^j$.

The compatibility condition between (\ref{lwzw}) and (\ref{hpartialwbjw}) gives
\be\label{hpartiallxj}
\hbar \frac{\partial L}{\partial x_j}=[B_j,L].
\ee
This is called the $\hbar$-dependent KP hierarchy, the $\hbar$-KP hierarchy for short. It is clear that when $\hbar\rightarrow 1$, it become the classical KP hierarchy.

Substituting (\ref{wM}) into (\ref{lwzw}), we get
\be
L=M\cdot (\hbar \partial)\cdot M^{-1}.
\ee
This equation gives the relations between $w_1,w_2,\cdots$ and $f_1,f_2,\cdots$. The compatibility condition (\ref{hpartiallxj}) shows that these unknown functions can be written in terms of a single function $\tau(x)$ by the following relation
\be
w=\frac{\tau (x-\hbar [z^{-1}])}{\tau(x)}e^{\sum_{j=1}^\infty\frac{x_j}{\hbar}z^j},
\ee
where
\[
[z^{-1}]=(\frac{1}{z}, \frac{1}{2z^2},\frac{1}{3z^3},\cdots).
\]
Then the $\hbar$-KP hierarchy is an infinite set of nonlinear differential equations in a function $\tau$ of infinitely many variables $x_1,x_2,\cdots$. As in the classical KP hierarchy, let
\be
u=2\hbar^2\partial^2 \text{log}\tau.
\ee
We get the first nonlinear differential equation in the $\hbar$-KP hierarchy
\be\label{hbarKP}
\frac{3}{4}\frac{\partial^2u}{\partial x_2^2}=\frac{\partial}{\partial x}\left(\frac{\partial u}{\partial x_3}-\frac{3}{2}u\frac{\partial u}{\partial x}-\frac{\hbar^2}{4}\frac{\partial^3u}{\partial x^3}\right).
\ee
This equation is called the $\hbar$-KP equation. When $\hbar\rightarrow 1$, it become the KP equation, and when $\hbar\rightarrow 0$, it become the dispersionless KP equation.
\section{Affine Yangian of ${\mathfrak{gl}}(1)$ and Jack polynomials}\label{sect3}
In this section, we recall the definitions of the affine Yangian of ${\mathfrak{gl}}(1)$ and the Jack polynomials. Then we show the properties of the Jack polynomials according to the affine Yangian of ${\mathfrak{gl}}(1)$.
\subsection{Affine Yangian of ${\mathfrak{gl}}(1)$}
We almost copy this section from that in \cite{deformedKPJack} since that is what we will use in the following of this paper. The affine Yangian $\mathcal{Y}$ of $\widehat{\mathfrak{gl}}(1)$ is an associative algebra with generators $e_j, f_j$ and $\psi_j$, $j = 0, 1, \ldots$ and the following relations\cite{Pro,Tsy}
\begin{eqnarray}
&&\left[ \psi_j, \psi_k \right] = 0,\\
&&\left[ e_{j+3}, e_k \right] - 3 \left[ e_{j+2}, e_{k+1} \right] + 3\left[ e_{j+1}, e_{k+2} \right] - \left[ e_j, e_{k+3} \right]\nonumber \\
&& \quad + \sigma_2 \left[ e_{j+1}, e_k \right] - \sigma_2 \left[ e_j, e_{k+1} \right] - \sigma_3 \left\{ e_j, e_k \right\} =0,\label{yangian1}\\
&&\left[ f_{j+3}, f_k \right] - 3 \left[ f_{j+2}, f_{k+1} \right] + 3\left[ f_{j+1}, f_{k+2} \right] - \left[ f_j, f_{k+3} \right] \nonumber\\
&& \quad + \sigma_2 \left[ f_{j+1}, f_k \right] - \sigma_2 \left[ f_j, f_{k+1} \right] + \sigma_3 \left\{ f_j, f_k \right\} =0, \label{yangian2}\\
&&\left[ e_j, f_k \right] = \psi_{j+k},\label{yangian3}\\
&& \left[ \psi_{j+3}, e_k \right] - 3 \left[ \psi_{j+2}, e_{k+1} \right] + 3\left[ \psi_{j+1}, e_{k+2} \right] - \left[ \psi_j, e_{k+3} \right]\nonumber \\
&& \quad + \sigma_2 \left[ \psi_{j+1}, e_k \right] - \sigma_2 \left[ \psi_j, e_{k+1} \right] - \sigma_3 \left\{ \psi_j, e_k \right\} =0,\label{yangian4}\\
&& \left[ \psi_{j+3}, f_k \right] - 3 \left[ \psi_{j+2}, f_{k+1} \right] + 3\left[ \psi_{j+1}, f_{k+2} \right] - \left[ \psi_j, f_{k+3} \right] \nonumber\\
&& \quad + \sigma_2 \left[ \psi_{j+1}, f_k \right] - \sigma_2 \left[ \psi_j, f_{k+1} \right] + \sigma_3 \left\{ \psi_j, f_k \right\} =0,\label{yangian5}
\end{eqnarray}
together with boundary conditions
\begin{eqnarray}
&&\left[ \psi_0, e_j \right]  = 0, \left[ \psi_1, e_j \right] = 0,  \left[ \psi_2, e_j \right]  = 2 e_j ,\label{yangian6}\\
&&\left[ \psi_0, f_j \right]  = 0,  \left[ \psi_1, f_j \right]  = 0,  \left[ \psi_2, f_j \right]  = -2f_j ,\label{yangian7}
\end{eqnarray}
and a generalization of Serre relations
\begin{eqnarray}
&&\mathrm{Sym}_{(j_1,j_2,j_3)} \left[ e_{j_1}, \left[ e_{j_2}, e_{j_3+1} \right] \right]  = 0, \label{yangian8} \\
&&\mathrm{Sym}_{(j_1,j_2,j_3)} \left[ f_{j_1}, \left[ f_{j_2}, f_{j_3+1} \right] \right]  = 0,\label{yangian9}
\end{eqnarray}
where $\mathrm{Sym}$ is the complete symmetrization over all indicated indices which include $6$ terms.

The notations $\sigma_2,\ \sigma_3$ in the definition of affine Yangian are functions of three complex numbers $h_1, h_2$ and $h_3$:
\begin{eqnarray}
\sigma_1 &=& h_1+h_2+h_3=0,\\
\sigma_2 &=& h_1 h_2 + h_1 h_3 + h_2 h_3,\\
 \sigma_3 &=& h_1 h_2 h_3.
\end{eqnarray}

The affine yangian $\mathcal{Y}$ has a representation on the plane partitions. A plane partition $\pi$ is a 2D Young diagram in the first quadrant of plane $xOy$ filled with non-negative integers that form nonincreasing rows and columns  \cite{FW, ORV}. The number in the position $(i,j)$ is denoted by $\pi_{i,j}$
\[
\left( \begin{array}{ccc}
\pi_{1,1} & \pi_{1,2} &\cdots \\
\pi_{2,1} & \pi_{2,2} &\cdots \\
\cdots & \cdots &\cdots \\
\end{array} \right).
\]
 The integers $\pi_{i,j}$ satisfy
\[
\pi_{i,j}\geq \pi_{i+1,j},\quad \pi_{i,j}\geq \pi_{i,j+1},\quad \lim_{i\rightarrow \infty}\pi_{i,j}=\lim_{j\rightarrow \infty}\pi_{i,j}=0
\]
for all integers $i,j\geq 0$. Piling $\pi_{i,j}$ cubes over position $(i, j)$ gives a 3D Young diagram. 3D Young diagrams arose naturally in the melting crystal model\cite{ORV,NT}. We always identify 3D Young diagrams with plane
partitions as explained above. For example, the 3D Young diagram
$\3Dtwoboxy$
can also be denoted by the plane partition
$(1,1)$.

As in our paper \cite{3DFermionYangian}, we use the following notations. For a 3D Young diagram $\pi$, the notation $\Box\in \pi^+$ means that this box is not in $\pi$ and can be added to $\pi$. Here ``can be added'' means that when this box is added, it is still a 3D Young diagram. The notation $\Box\in \pi^-$ means that this box is in $\pi$ and can be removed from $\pi$. Here ``can be removed" means that when this box is removed, it is still a 3D Young diagram. For a box $\Box$, we let
\begin{equation}\label{epsilonbox}
h_\Box=h_1y_\Box+h_2x_\Box+h_3z_\Box,
\end{equation}
where $(x_\Box,y_\Box,z_\Box)$ is the coordinate of box $\Box$ in coordinate system $O-xyz$. Here we use the order $y_\Box,x_\Box,z_\Box$ to match that in paper \cite{Pro}.

Following \cite{Pro,Tsy}, we introduce the generating functions:
\begin{eqnarray}
e(u)&=&\sum_{j=0}^{\infty} \frac{e_j}{u^{j+1}},\nonumber\\
f(u)&=&\sum_{j=0}^{\infty} \frac{f_j}{u^{j+1}},\\
\psi(u)&=& 1 + \sigma_3 \sum_{j=0}^{\infty} \frac{\psi_j}{u^{j+1}},\nonumber
\end{eqnarray}
where $u$ is a parameter.
Introduce
\begin{equation}\label{psi0}
\psi_0(u)=\frac{u+\sigma_3\psi_0}{u}
\end{equation}
and
\begin{eqnarray} \label{dfnvarphi}
\varphi(u)=\frac{(u+h_1)(u+h_2)(u+h_3)}{(u-h_1)(u-h_2)(u-h_3)}.
\end{eqnarray}
For a 3D Young diagram $\pi$, define $\psi_\pi(u)$ by
\begin{eqnarray}\label{psipiu}
\psi_\pi(u)=\psi_0(u)\prod_{\Box\in\pi} \varphi(u-h_\Box).
\end{eqnarray}
In the following, we recall the representation of the affine Yangian on 3D Young diagrams as in paper \cite{Pro} by making a slight change. The representation of affine Yangian on 3D Young diagrams is given by
\begin{eqnarray}
\psi(u)|\pi\rangle&=&\psi_\pi(u)|\pi\rangle,\\
e(u)|\pi\rangle&=&\sum_{\Box\in \pi^+}\frac{E(\pi\rightarrow\pi+\Box)}{u-h_\Box}|\pi+\Box\rangle,\label{eupi}\\
f(u)|\pi\rangle&=&\sum_{\Box\in \pi^-}\frac{F(\pi\rightarrow\pi-\Box)}{u-h_\Box}|\pi-\Box\rangle\label{fupi}
\end{eqnarray}
where $|\pi\rangle$ means the state characterized by the 3D Young diagram $\pi$ and the coefficients
\begin{equation}\label{efpi}
E(\pi \rightarrow \pi+\square)=-F(\pi+\square \rightarrow \pi)=\sqrt{\frac{1}{\sigma_3} \operatorname{res}_{u \rightarrow h_{\square}} \psi_\pi(u)}
\end{equation}
 Equations (\ref{eupi}) and (\ref{fupi}) mean generators $e_j,\ f_j$ acting on the 3D Young diagram $\pi$ by
\begin{equation}
\begin{aligned}
e_j|\pi\rangle &=\sum_{\square \in \pi^{+}} h_{\square}^j E(\pi \rightarrow \pi+\square)|\pi+\square\rangle,
\end{aligned}
\end{equation}
\begin{equation}
\begin{aligned}
f_j|\pi\rangle &=\sum h_{\square}^j F(\pi \rightarrow \pi-\square)|\pi-\square\rangle .
\end{aligned}
\end{equation}

As in our paper \cite{3DFermionYangian}, we use the following notations.
3D Young diagram $\pi$ may have many ways to get by adding box, for example, there are two ways to get 3D Young diagram $(2,1)$, which are
\beaa
&&(1)\rightarrow (1,1)\rightarrow (2,1),\\
&&(1)\rightarrow (2)\rightarrow (2,1),
\eeaa
We denote the state corresponding to $(2,1)$ in the first equation of the two equations above by $|(2,1)\rangle_{h_1,h_3}$, and that in the second equation above by $|(2,1)\rangle_{h_3,h_1}$. We explain the subscripts: let $h_\Box=x_\Box h_2+ y_\Box h_1+z_\Box h_3$ with $h_1+h_2+h_3=0$, we use $h_\Box$-position to represent position $(x_\Box, y_\Box, z_\Box)$ in coordinate system $O-xyz$. The notation ``$h_1,h_3$'' means adding one box to $\Box$ in $h_1$-position  first, then adding one box in $h_3$-position. Even though $h_1 $-position is not unique, for example, $h_1 $-position can be the positions $(1,2,1),(2,3,2),\cdots$ since $h_1+h_2+h_3=0$, but it is unique if we want to get a new 3D Young diagram after adding this box. Therefore, we can read the notation $|\left(\begin{array}{ccc} 1 &1\\ 1 &\end{array}\right)\rangle_{h_1,h_2}$, which means the 3D Young diagram $|\left(\begin{array}{ccc} 1 &1\\ 1 &\end{array}\right)\rangle$ is obtained from $\Box $ by adding one box in $h_1$-position first, then adding one box in $h_2$-position. When there is no confusion, we will omit the subscripts.

The state corresponding to 3D Young diagram is related to its growth process, this is because we denote $E(\pi\rightarrow\pi+\Box)|\pi+\Box\rangle$ by $|\pi+\Box\rangle$ the the 3D Young diagram representation of affine Yangian of $\mathfrak{gl}(1)$. For example,
\beaa
|(2,1)\rangle_{h_1,h_3}&=& E((1)\rightarrow (1,1))E((1,1)\rightarrow (2,1))|(2,1)\rangle,\\
|(2,1)\rangle_{h_3,h_1}&=& E((1)\rightarrow (2))E((2)\rightarrow (2,1))|(2,1)\rangle,
\eeaa
then, $|(2,1)\rangle_{h_1,h_3}=\varphi(h_3-h_1)|(2,1)\rangle_{h_3,h_1}$.

In the following subsection, we will discuss the Jack polynomials $\tilde{J}_\lambda(x)$, where we treat 2D Young diagrams as the special cases of 3D Young diagrams which have one layer in $z$-axis direction. The symmetric functions $\tilde{J}_\lambda(x)$ in the next subsection behave as the special case $\psi_0=1,\ h_1=\sqrt{\alpha}, \ h_2=-\sqrt{\alpha}^{-1}$ of the 3D Young diagrams in this subsection.
\subsection{The Jack polynomials}
The Jack polynomials we discussed here are denoted by $\tilde{J}_\lambda$ or $\tilde{\tilde{J}}_\lambda$ since it is well known that the notations $J_\lambda$ are used in \cite{Mac}. The Jack polynomials $\tilde{J}_\lambda$ \cite{JackYangian,deformedKPJack} equal the Jack polynomials $P_\lambda^{\alpha}$ (defined in \cite{Mac}) multiplied by a constant. We introduce the Jack polynomials $\tilde{J}_\lambda$ since they behave the same as the Young diagrams in the last section. For example, it can be checked that $\langle \tilde{J}_\lambda,\tilde{J}_\mu\rangle$ in the following equals $\langle \lambda,\mu\rangle$ defined in the last section.

Let ${ p}=(p_1,p_2,\cdots)$, the Jack polynomials $\tilde{J}_\lambda$ are defined by\cite{JackYangian}
\begin{equation}
\tilde{J}_{ \lambda} :=\frac{ B_{\lambda}       }{ A_{\lambda}} P_{\lambda }^{\alpha }
\end{equation}
where
\begin{eqnarray}
A_{\lambda} &=&(\sqrt{\alpha }) ^{\lambda_{1} -\lambda_{2}-\dots -\lambda_{l}}\cdot \prod_{j=1}^{l-1}  \prod_{i=0}^{\lambda_{l}-1} \left [ \left ( \lambda_{j}-i \right )\alpha +l-j   \right ] \cdot  \prod_{i=1}^{l-1}  \prod_{j=1}^{\lambda_{i+1} } \left [ \left ( \lambda_{i} -j \right )\alpha +1   \right ]\nn\\
&&  \cdot\prod_{k=3}^{{l} }   \prod_{j=0}^{\lambda_{k-1} -\lambda_{k}-1}  \prod_{i=1}^{k-2} \left [ \left ( \lambda_{i} -\lambda_{k} -j \right )\alpha +k-i-1   \right ],\\
B_{\lambda} &=&\prod_{i=1}^{l-1}  \prod_{j=0}^{\lambda_{i+1}-1} \left ( \lambda_{i}  -j \right )\cdot   \prod_{i=1}^{l-1}  \prod_{j=1}^{\lambda_{l}}\left [ \left ( \lambda_{i}  -j \right )\alpha +l-i+1  \right ] \nn\\
&&\cdot \prod_{k=3}^{{l} }   \prod_{j=1}^{\lambda_{k-1} -\lambda_{k}}  \prod_{i=1}^{k-2} \left [ \left ( \lambda_{i} -\lambda_{k} -j \right )\alpha +k-i  \right ].
\end{eqnarray}
for 2D Young diagram $\lambda=(\lambda_1,\lambda_2,\cdots, \lambda_l)$.

We change $p_n$ in \cite{Mac} to $\sqrt{\alpha}p_n$, then $p_n$ in the following satisfies $\langle p_n,p_n\rangle=n$.
Then Jack polynomials $\tilde{J}_{(n)}$ defined above satisfy
\be
\exp\left(\sum_{n\geq 1}\frac{p_n}{n\sqrt{\alpha}}z^n\right)=\sum_{n\geq 0}\frac{1}{\langle \tilde{J}_{(n)},\tilde{J}_{(n)}\rangle}\frac{1}{\sqrt{\alpha}^n}\tilde{J}_{(n)}z^n.\label{x+zyn}
\ee
Define the operator $\hat{\tilde J}_{(n)}$ by
\be
\exp\left(\sum_{n\geq 1}\frac{\text{ad}_{e_1}^{n-1}e_0}{n!\sqrt{\alpha}}z^n\right)=\sum_{n\geq 0}\frac{1}{\langle \tilde{J}_{(n)},\tilde{J}_{(n)}\rangle}\frac{1}{\sqrt{\alpha}^n}\hat{\tilde{J}}_{(n)}z^n.\label{y_{(n)}}
\ee
The Pieri formula $\tilde{J}_{(n)}\tilde{J}_\lambda$ is defined by
\be\label{pieri}
\tilde{J}_{(n)}\tilde{J}_\lambda:=\hat{\tilde{J}}_{(n)}\cdot \tilde{J}_\lambda.
\ee
Note that the actions of the generators $e_k,\ f_k,\ \psi_k $ of affine Yangian of ${\mathfrak{gl}}(1)$ on $\tilde{J}_\lambda$ are the same with that on $\lambda$. The expressions of $\tilde{J}_{\lambda}$ for all Young diagrams $\lambda$ can be determined by (\ref{y_{(n)}}) and (\ref{pieri}). Note that $\tilde{J}_\lambda$ can not be expressed as the determinant of $\tilde{J}_{(n)}$, while Schur functions $S_\lambda$ can be expressed as the determinant of $S_{(n)}$. Then we define $\tilde{\tilde{J}}_\lambda$ by
\be
\tilde{\tilde J}_\lambda:=\det \left(\frac{1}{\langle \tilde{J}_{\lambda_j-i+j},\tilde{J}_{\lambda_j-i+j}\rangle\sqrt{\alpha}^{\lambda_j-i+j}}\tilde{J}_{\lambda_j-i+j}\right)_{1\leq i,j\leq k},
\ee
which is slightly different from that in \cite{deformedKPJack}.
The transition matrix $M=(M_{\lambda\mu})$ from the set $\{\tilde{J}_\lambda\}$ to the set $\tilde{\tilde{J}}_\lambda$ is upper triangular\cite{deformedKPJack}, with the elements $M_{\lambda\lambda}$ in the diagonal equal
\be
\frac{1}{\sqrt{\alpha}^{|\lambda|}}\frac{1}{\langle \tilde{J}_{\lambda},\tilde{J}_{\lambda}\rangle}\prod_{j=2}^k\frac{(j\alpha)^{\lambda_j}}{(1+(j-1)\alpha)^{\lambda_j}}.
\ee

Introduce Bosons $a_n, \ n\in\Z,\ n\neq 0$ with the relations
\be\label{bosonsan}
[a_n,a_m]=n\delta_{n+m,0}.
\ee
In fact, for $n>0$,\cite{WBCW}
\be
a_{-n}=\frac{1}{(n-1)!}\text{ad}_{e_1}^{n-1}e_0,\ \ a_{n}=-\frac{1}{(n-1)!}\text{ad}_{f_1}^{n-1}f_0.
\ee
On Jack polynomials $\tilde{J}_\lambda$, the Bosons $a_n$ can be represented as
\be
a_{-n}=p_n,\ \ a_n=n\partial_{p_n}.
\ee

In the following of this subsection, we discuss the properties related to the Jack polynomials $\tilde{J}_\lambda(p)$ and $\tilde{\tilde{J}}_\lambda(p)$. Let $\partial_p=(\partial_{p_1},2\partial_{p_2},3\partial_{p_3},\cdots)$. From (\ref{x+zyn}), we have
\be
\exp\left(\sum_{n=1}^\infty\frac{\partial_{p_n}}{\sqrt{\alpha}}z^n\right)=\sum_{n\geq 0}\frac{1}{\langle \tilde{J}_{(n)},\tilde{J}_{(n)}\rangle \sqrt{\alpha}^n} \tilde{J}_n(\partial_p) z^n.
\ee
From
\bea
&&\exp\left(\sum_{n=1}^\infty\frac{\partial_{p_n}}{\sqrt{\alpha}}z^n\right)\exp\left(\sum_{n\geq 1}\frac{p_n}{n\sqrt{\alpha}}w^n\right)\nn\\
&=&\frac{1}{(1-zw)^{\frac{1}{\alpha}}}\exp\left(\sum_{n\geq 1}\frac{p_n}{n\sqrt{\alpha}}w^n\right)\exp\left(\sum_{n=1}^\infty\frac{\partial_{p_n}}{\sqrt{\alpha}}z^n\right),
\eea
we get that the operators $\tilde{J}_{(n)}(\partial_{p})$ and $\tilde{J}_{(m)}(p)$ satisfy
\bea
&&\frac{1}{\langle \tilde{J}_{(n)},\tilde{J}_{(n)}\rangle \sqrt{\alpha}^n}\tilde{J}_{(n)} (\partial_p)\frac{1}{\langle \tilde{J}_{(m)},\tilde{J}_{(m)}\rangle \sqrt{\alpha}^m}\tilde{J}_{(m)} (p)\\
&=&\sum_{k\geq 0}\left(\begin{array}{cc}-1/{\alpha}\\ k\end{array}\right)(-1)^k\frac{1}{\langle \tilde{J}_{(m-k)},\tilde{J}_{(m-k)}\rangle \sqrt{\alpha}^{m-k}}\tilde{J}_{(m-k)} (p)\frac{1}{\langle \tilde{J}_{(n-k)},\tilde{J}_{(n-k)}\rangle \sqrt{\alpha}^{n-k}}\tilde{J}_{(n-k)}(\partial_p),\nn
\eea
and the operators $\tilde{J}_{(n)}(\partial_{p})$ acting on the polynomials $\tilde{J}_{(m)}(p)$ equals
\bea
&&\frac{1}{\langle \tilde{J}_{(n)},\tilde{J}_{(n)}\rangle \sqrt{\alpha}^n}\tilde{J}_{(n)}(\partial_p)\frac{1}{\langle \tilde{J}_{(m)},\tilde{J}_{(m)}\rangle \sqrt{\alpha}^m}\tilde{J}_{(m)} (p)\nn\\
&=&\left(\begin{array}{cc}-1/{\alpha}\\ n\end{array}\right)(-1)^n\frac{1}{\langle \tilde{J}_{(m-n)},\tilde{J}_{(m-n)}\rangle \sqrt{\alpha}^{m-n}}\tilde{J}_{(m-n)} (p),
\eea
where we let $\tilde{J}_{(n)}=0$ unless $n\geq 0$.

The polynomials $\tilde{J}_{1^n}(p)$ satisfy
\be
\exp\left(-\sum_{n\geq 1}\frac{p_n}{n}\sqrt{\alpha}z^{n}\right)=\sum_{n=0}^\infty\frac{(-1)^n\sqrt{\alpha}^n}{\langle \tilde{J}_{(1^n)},\tilde{J}_{(1^n)}\rangle }\tilde{J}_{(1^n)}(p)z^n.
\ee
Then
\be
\exp\left(-\sum_{n\geq 1}\partial_{p_n}\sqrt{\alpha}z^{n}\right)=\sum_{n=0}^\infty\frac{(-1)^n\sqrt{\alpha}^n}{\langle \tilde{J}_{(1^n)},\tilde{J}_{(1^n)}\rangle }\tilde{J}_{(1^n)}(\partial p)z^n.
\ee
From
\bea
&&\exp\left(-\sum_{n\geq 1}\partial_{p_n}\sqrt{\alpha}z^{n}\right)\exp\left(-\sum_{n\geq 1}\frac{p_n}{n}\sqrt{\alpha}w^{n}\right)\nn\\
&=&\frac{1}{(1-zw)^\alpha}\exp\left(-\sum_{n\geq 1}\frac{p_n}{n}\sqrt{\alpha}w^{n}\right)\exp\left(-\sum_{n\geq 1}\partial_{p_n}\sqrt{\alpha}z^{n}\right),
\eea
we obtain that the operators $\tilde{J}_{1^n}(\partial_p)$ and $\tilde{J}_{1^m}(p)$ satisfy
\bea
&&\frac{(-1)^n\sqrt{\alpha}^n}{\langle \tilde{J}_{(1^n)},\tilde{J}_{(1^n)}\rangle}\tilde{J}_{(1^n)} (\partial_p)\frac{(-1)^m\sqrt{\alpha}^m}{\langle \tilde{J}_{(1^m)},\tilde{J}_{(1^m)}\rangle }\tilde{J}_{(1^m)} (p)\\
&=&\sum_{k\geq 0}\left(\begin{array}{cc}-{\alpha}\\ k\end{array}\right)(-1)^k\frac{(-1)^{m-k}\sqrt{\alpha}^{m-k}}{\langle \tilde{J}_{(m-k)},\tilde{J}_{(m-k)}\rangle }\tilde{J}_{(m-k)} (p)\frac{(-1)^{n-k}\sqrt{\alpha}^{n-k}}{\langle \tilde{J}_{(1^{n-k})},\tilde{J}_{(1^{n-k})}\rangle}\tilde{J}_{(1^{n-k})}(\partial_p),\nn
\eea
and the operators $\tilde{J}_{(1^n)}(\partial_{p})$ acting on the polynomials $\tilde{J}_{(1^m)}(p)$ equals
\bea
&&\frac{\sqrt{\alpha}^n}{\langle \tilde{J}_{(1^n)},\tilde{J}_{(1^n)}\rangle }\tilde{J}_{(1^n)}(\partial_p)\frac{\sqrt{\alpha}^m}{\langle \tilde{J}_{(1^m)},\tilde{J}_{(1^m)}\rangle \sqrt{\alpha}^m}\tilde{J}_{(m)} (p)\nn\\
&=&\left(\begin{array}{cc}-{\alpha}\\ n\end{array}\right)(-1)^n\frac{\sqrt{\alpha}^{m-n}}{\langle \tilde{J}_{(1^{m-n})},\tilde{J}_{(1^{m-n})}\rangle}\tilde{J}_{(1^{m-n}))} (p),
\eea
where we let $\tilde{J}_{(1^n)}=0$ unless $n\geq 0$.
From
\bea
&&\exp\left(-\sum_{n\geq 1}{\partial_{p_n}}\sqrt{\alpha}z^{n}\right)\exp\left(\sum_{n\geq 1}\frac{p_n}{\sqrt{\alpha}}w^{n}\right)\\
&=&{(1-zw)}\exp\left(\sum_{n\geq 1}\frac{p_n}{\sqrt{\alpha}}w^{n}\right)\exp\left(-\sum_{n\geq 1}{\partial_{p_n}}\sqrt{\alpha}z^{n}\right),
\eea
we obtain the operators $\tilde{J}_{(1^n)}(\partial_p)$ and $\tilde{J}_{(m)}$ satisfy
\bea
&&\frac{1}{\langle \tilde{J}_{(1^n)},\tilde{J}_{(1^n)}\rangle}\tilde{J}_{(1^n)}(\partial_p)\frac{1}{\langle \tilde{J}_{(m)},\tilde{J}_{(m)}\rangle}\tilde{J}_{(m)}(p)\nn\\
&=&\frac{1}{\langle \tilde{J}_{(m)},\tilde{J}_{(m)}\rangle}\tilde{J}_{(m)}(p)\frac{1}{\langle \tilde{J}_{(1^n)},\tilde{J}_{(1^n)}\rangle}\tilde{J}_{(1^n)}(\partial_p)\nn\\
&+&\frac{1}{\langle \tilde{J}_{(m-1)},\tilde{J}_{(m-1)}\rangle}\tilde{J}_{(m-1)}(p)\frac{1}{\langle \tilde{J}_{(1^{n-1})},\tilde{J}_{(1^{n-1})}\rangle}\tilde{J}_{(1^{n-1})}(\partial_p).
\eea
From
\bea
&&\exp\left(-\sum_{n\geq 1}{\partial_{p_n}}\sqrt{\alpha}z^{n}\right)\exp\left(-\sum_{n\geq 1}\frac{p_n}{\sqrt{\alpha}}w^{n}\right)\\
&=&\frac{1}{1-zw}\exp\left(\sum_{n\geq 1}\frac{p_n}{\sqrt{\alpha}}w^{n}\right)\exp\left(-\sum_{n\geq 1}{\partial_{p_n}}\sqrt{\alpha}z^{n}\right),
\eea
we obtain the operators $\tilde{J}_{(1^n)}(\partial_p)$ and $\tilde{J}_{(m)}(p)$ satisfy
\bea
&&\frac{(-1)^n\sqrt{\alpha}^n}{\langle \tilde{J}_{(1^n)},\tilde{J}_{(1^n)}\rangle}\tilde{J}_{(1^n)}(\partial_p)\frac{1}{\langle \tilde{J}_{(m)},\tilde{J}_{(m)}\rangle\sqrt{\alpha}^m}\tilde{J}_{(m)}(p)\nn\\
&=&\sum_{k\geq 0}\frac{1}{\langle \tilde{J}_{(m-k)},\tilde{J}_{(m-k)}\rangle\sqrt{\alpha}^{m-k}}\tilde{J}_{(m-k)}(p)\frac{(-1)^{n-k}\sqrt{\alpha}^{n-k}}{\langle \tilde{J}_{(1^{n-k})},\tilde{J}_{(1^{n-k})}\rangle}\tilde{J}_{(1^{n-k})}(\partial_p).
\eea
Introduce the vertex operators
\bea
X_+(z)&=&\sum_{n\in\Z}X_n^+ z^n=\exp\left(\sum_{n\geq 1}\frac{p_n}{n\sqrt{\alpha}}z^{n}\right)\exp\left(-\sum_{n\geq 1}{\partial_{p_n}}\sqrt{\alpha}z^{-n}\right)\label{vertexv+zpn},\\
X_-(z)&=&\sum_{n\in\Z}X_n^- z^n=\exp\left(-\sum_{n\geq 1}\frac{p_n}{n\sqrt{\alpha}}z^{n}\right)\exp\left(\sum_{n\geq 1}{\partial_{p_n}}\sqrt{\alpha}z^{-n}\right).\label{vertexv-zpn}
\eea
The Jack polynomials $\tilde{\tilde{J}}_\lambda(p)$ have the vertex operator realization\cite{deformedKPJack}
\be
\tilde{\tilde J}_\lambda=X_{\lambda_1}^+X_{\lambda_2}^+\cdots X_{\lambda_k}^+ \cdot 1
\ee
for $\lambda=(\lambda_1,\lambda_2,\cdots,\lambda_k)$.

From (\ref{x+zyn}), we know that
\be
\frac{1}{\langle \tilde{J}_{(n)},\tilde{J}_{(n)}\rangle}\frac{1}{\sqrt{\alpha}^n}\tilde{J}_{(n)}=S_{(n)}\left(\frac{p}{\sqrt{\alpha}}\right).
\ee
Then from the definition of $\tilde{\tilde{J}}_\lambda(p)$, we have
\be
\tilde{\tilde{J}}_\lambda(p)=S_{\lambda}\left(\frac{p}{\sqrt{\alpha}}\right).
\ee
In \cite{deformedKPJack}, an integrable hierarchy is defined to be the bilinear relations:
\be
\sum_{m+n=-1} X_{m}^-\tau\otimes X_n^+\tau=0,\label{skp}
\ee
where $\tau=\tau (x)$ is an unknown function. In the next section, we will show that this hierarchy is exactly the $\hbar$-KP hierarchy defined in the last section.

The set $\{\tilde{J}_\lambda\}$ is an orthogonal basis, but $\{\tilde{\tilde{J}}_\lambda\}$ is not. From
\[
\langle \tilde{J}_{(n)},\tilde{J}_{(n)}\rangle=\prod_{k=1}^{n}\frac{k}{1+(k-1)\alpha},
\]
we have
\be
\langle S_{(n)}\left(\frac{p}{\sqrt{\alpha}}\right),S_{(n)}\left(\frac{p}{\sqrt{\alpha}}\right)\rangle =\prod_{k=1}^{n}\frac{1+(k-1)\alpha}{k\alpha}.
\ee
From
\bea
S_{(n,1)}\left(\frac{p}{\sqrt{\alpha}}\right)&=&\frac{1}{\sqrt{\alpha}^{n-1}}\frac{1}{\langle J_{(n,1)},J_{(n,1)}\rangle}\frac{2}{1+\alpha}\tilde{J}_{(n,1)}+\frac{n(1-\alpha)}{1+n\alpha}S_{(n+1)}\left(\frac{p}{\sqrt{\alpha}}\right),
\eea
where $(n,1)$ is the Young diagram obtained $(1,1)$ by adding $(n-1)$ box,
we have
\bea
\langle S_{(n+1)}\left(\frac{p}{\sqrt{\alpha}}\right),S_{(n,1)}\left(\frac{p}{\sqrt{\alpha}}\right)\rangle =\frac{n(1-\alpha)}{1+n\alpha}\prod_{k=1}^{n+1}\frac{1+(k-1)\alpha}{k\alpha},
\eea
which shows that $S_{(n+1)}\left(\frac{p}{\sqrt{\alpha}}\right)$ and $S_{(n,1)}\left(\frac{p}{\sqrt{\alpha}}\right)$ is not orthogonal. The set $\{S_{\lambda}\left(\frac{p}{\sqrt{\alpha}}\right)\}$ is still a basis since the transition matrix from the set $\{\tilde{J}_\lambda\}$ to $\{\tilde{\tilde{J}}_\lambda\}$ is upper triangular in the sense of Young diagram's reverse lexicographical order. For example,
\beaa
S_{(n,1)}\left(\frac{p}{\sqrt{\alpha}}\right)&=&\frac{1}{\langle \tilde{J}_{(1)},\tilde{J}_{(1)}\rangle\sqrt{\alpha}}\tilde{J}_{(1)}\frac{1}{\langle \tilde{J}_{(n)},\tilde{J}_{(n)}\rangle\sqrt{\alpha}^{n}}\tilde{J}_{(n)}-\frac{1}{\langle \tilde{J}_{(n+1)},\tilde{J}_{(n+1)}\rangle\sqrt{\alpha}^{n+1}}\tilde{J}_{(n+1)}\\
&=&\frac{1}{\langle \tilde{J}_{(n)},\tilde{J}_{(n)}\rangle\sqrt{\alpha}^{n+1}}\tilde{J}_{(n,1)}+\frac{1}{\sqrt{\alpha}^{n+1}}\left(\frac{1}{\langle \tilde{J}_{(n)},\tilde{J}_{(n)}}-\frac{1}{\langle \tilde{J}_{(n+1)},\tilde{J}_{(n+1)}}\right)\tilde{J}_{n+1},
\eeaa
and
\beaa
S_{(n,2)}\left(\frac{p}{\sqrt{\alpha}}\right)&=&\frac{1}{\langle \tilde{J}_{(2)},\tilde{J}_{(2)}\rangle\sqrt{\alpha}^2}\tilde{J}_{(2)}\frac{1}{\langle \tilde{J}_{(n)},\tilde{J}_{(n)}\rangle\sqrt{\alpha}^{n}}\tilde{J}_{(n)}\\
&&-\frac{1}{\langle \tilde{J}_{(1)},\tilde{J}_{(1)}\rangle\sqrt{\alpha}}\tilde{J}_{(1)}\frac{1}{\langle \tilde{J}_{(n+1)},\tilde{J}_{(n+1)}\rangle\sqrt{\alpha}^{n+1}}\tilde{J}_{(n+1)}.
\eeaa

We show the actions of the Bosons on $S_{\lambda}\left(\frac{p}{\sqrt{\alpha}}\right)$.
\be
a_{-1}S_{\lambda}\left(\frac{p}{\sqrt{\alpha}}\right)=p_1S_{\lambda}\left(\frac{p}{\sqrt{\alpha}}\right)=\sqrt{\alpha}\sum_{\Box\in\lambda^+}S_{\lambda+\Box}\left(\frac{p}{\sqrt{\alpha}}\right),
\ee
\be
a_{-2}S_{\lambda}\left(\frac{p}{\sqrt{\alpha}}\right)=\sqrt{\alpha}\sum_{\begin{tikzpicture}
\draw (0.25,0) -- (0.25,0.25) [-];
\draw (0,0)rectangle(0.5,0.25);
\end{tikzpicture}\in\lambda^+}S_{\lambda+\begin{tikzpicture}
\draw (0.25,0) -- (0.25,0.25) [-];
\draw (0,0)rectangle(0.5,0.25);
\end{tikzpicture}}\left(\frac{p}{\sqrt{\alpha}}\right)-\sqrt{\alpha}\sum_{\begin{tikzpicture}
\draw (0,0.25) -- (0.25,0.25) [-];
\draw (0,0)rectangle(0.25,0.5);
\end{tikzpicture}\in\lambda^+}S_{\lambda+\begin{tikzpicture}
\draw (0,0.25) -- (0.25,0.25) [-];
\draw (0,0)rectangle(0.25,0.5);
\end{tikzpicture}}\left(\frac{p}{\sqrt{\alpha}}\right).
\ee
In fact, for $n>0$, the Bosons $a_{-n}$ acting on $S_{\lambda}\left(\frac{p}{\sqrt{\alpha}}\right)$ here equals $a_{-n}$ acting on $S_\lambda(p)$ in the KP hierarchy \cite{Mac} multiplied by $\sqrt{\alpha}$, while the Bosons $a_{n}$ acting on $S_{\lambda}\left(\frac{p}{\sqrt{\alpha}}\right)$ here equals $a_{n}$ acting on $S_\lambda(p)$ in the KP hierarchy \cite{Mac} multiplied by $1/\sqrt{\alpha}$.
\section{The Hirota equation and vertex operators}\label{sect4}
In this section, we will show that the $\hbar$-dependent KP hierarchy is exactly the integrable hierarchy defined in (\ref{skp}). The calculation is similar to that of KP hierarchy in \cite{Mac}.

Let $x=(x_1,x_2,\cdots)$ and $y=(y_1,y_2,\cdots)$. The Hirota derivative $D_{x_k}^j$ is defined to be\cite{Mac}
\be
D_{x_k}^jf(x)\cdot g(x)=\partial_{y_k}^jf(x+y)g(x-y)|_{y=0}
\ee
The $\hbar$-KP equation (\ref{hbarKP}) becomes the following Hirota equation
\be\label{hbareqHirota}
\hbar^2 D_{x_1}^4 \tau(x)\cdot \tau(x)-4D_{x_1}D_{x_3}\tau(x)\cdot \tau(x)+3D_{x_2}^2\tau(x)\cdot \tau(x)=0.
\ee
Set
\[
P(k_1,k_2,k_3)=\hbar^2 k_1^4+3k_2^2-4k_1k_3.
\]
The solutions of the equation $P(k_1,k_2,k_3)=0$ are
\[
(k_1,k_2,k_3)=(\frac{1}{\hbar}p-\frac{1}{\hbar}q,
\frac{1}{\hbar}p^2-\frac{1}{\hbar}q^2,
\frac{1}{\hbar}p^3-\frac{1}{\hbar}q^3)
\]
for any $z_1,z_2$. Take two solutions
\beaa
(k_1,k_2,k_3)&=&(\frac{1}{\hbar}p_1-\frac{1}{\hbar}q_1,
\frac{1}{\hbar}p_1^2-\frac{1}{\hbar}q_1^2,
\frac{1}{\hbar}p_1^3-\frac{1}{\hbar}q_1^3),\\
(k_1',k_2',k_3')&=&(\frac{1}{\hbar}p_2-\frac{1}{\hbar}q_2,
\frac{1}{\hbar}p_2^2-\frac{1}{\hbar}q_2^2,
\frac{1}{\hbar}p_2^3-\frac{1}{\hbar}q_2^3),
\eeaa
we have
\[
-\frac{P(k_1-k_1',k_2-k_2',k_3-k_3')}{P(k_1+k_1',k_2+k_2',k_3+k_3')}=\frac{(p_1-p_2)(q_1-q_2)}{(p_1-q_2)(q_1-p_2)}.
\]
Suppose
\beaa
\xi_i&=&\sum_{j=1}^\infty(p_i^j-q_i^j)\frac{x_j}{\hbar},\\
a_{ii'}&=&\frac{(p_i-p_{i'})(q_i-q_{i'})}{(p_i-q_{i'})(q_i-p_{i'})},
\eeaa
then for $I=\{1,2,\cdots,n\}$,
\be
\tau=\sum_{J\subset I}\left(\prod_{i\in J}c_i\right)\left(\prod_{i,i'\in J,i<i'}a_{ii'}\right)\exp\left(\sum_{i\in J}\xi_i\right)
\ee
gives the $n$-soliton solution of $\hbar$-KP hierarchy. Introduce
the vertex operator
\be
X(p,q)=\exp\left(\sum_{j=1}^\infty(p^j-q^j)\frac{x_j}{\hbar}\right)
\exp\left(-\sum_{j=1}^\infty\frac{\hbar}{j}(p^{-j}-q^{-j})\partial_{x_j}\right),
\ee
then the $n$-soliton solution $\tau$ above can be written as
\be
\tau=e^{c_1X(p_1,q_1)}\cdots e^{c_nX(p_n,q_n)}\cdot 1.
\ee

The tau functions satisfy the following bilinear identity. For any $x$ and $x'$, let
\[
\xi=\sum_{j=1}^\infty\frac{x_j}{\hbar}z^j,\ \ \xi'=\sum_{j=1}^\infty\frac{x_j'}{\hbar}z^j.
\]
the bilinear identity holds:
\be\label{bilinearid}
\oint\frac{dz}{2\pi \sqrt{-1}}e^{\xi-\xi'}\tau(x-\hbar [z^{-1}])\tau(x'+\hbar [z^{-1}])=0.
\ee

Introduce
\be
w^*(x,z)=\frac{\tau(x+\hbar [z^{-1}])}{\tau(x)}e^{-\sum_{j=1}^\infty\frac{x_j}{\hbar}z^j},
\ee
it has the following form
\[
w^*(x,z)=e^{-\sum_{j=1}^\infty\frac{x_j}{\hbar}z^j}\left(1+\sum_{j=1}^\infty\frac{w_j^*}{z^j}\right).
\]
Then the bilinear identity (\ref{bilinearid}) becomes
\be
\oint \frac{dz}{2\pi\sqrt{-1}}w(x,z)w^*(x',z)=0.
\ee
From this relation, the linear system (\ref{hpartialwbjw}) can be obtained. Let
\[
Q=\hbar \partial_{x_j}-(L^j)_+,
\ \
\tilde{w}(x,k)=Qw(x,k).
\]
Then $\tilde{w}(x,k)$ has the following form
\[
\tilde{w}(x,k)=e^{\sum_{j=1}^\infty\frac{x_j}{\hbar}z^j}\left(\sum_{j=1}^\infty\frac{\tilde{w}_j}{z^j}\right)
\]
and satisfy
\[
\oint \frac{dz}{2\pi\sqrt{-1}}\tilde{w}(x,z)w^*(x',z)=0,
\]
which show $\tilde{w}_j=0$ for $j=1,2,\cdots$. Therefore, $Qw(x,z)=0$, that is (\ref{hpartialwbjw}) is obtained.

Let $x_n=p_n/n$ and $\alpha=\hbar^2$, we see that the bilinear identity (\ref{bilinearid}) is the same with (\ref{skp}). Then the polynomials $\tilde{\tilde{J}}_\lambda(p)=S_\lambda\left(\frac{p}{\sqrt{\alpha}}\right)$ are all the solutions of the $\hbar$-KP hierarchy.

We have calculated that the first equation in the $\hbar$-KP hierarchy has the form\cite{deformedKPJack}
\bea
&&\frac{4}{(1+\alpha)^2}\frac{1}{\langle \tilde{J}_{(2,2)},\tilde{J}_{(2,2)}\rangle}\left(\tilde{J}_{(2,2)}(\partial x)\tau\right)\cdot \tau\nn\\
&+&\frac{6(\alpha-1)}{(2+\alpha)(3+\alpha)}\frac{1}{\langle \tilde{J}_{(2,1,1)},\tilde{J}_{(2,1,1)}\rangle}\left(\tilde{J}_{(2,1,1)}(\partial x)\tau\right)\cdot \tau\nn\\
&+&\frac{2(\alpha-1)\alpha}{(2+\alpha)(3+\alpha)}\frac{1}{\langle \tilde{J}_{(1^4)},\tilde{J}_{(1^4)}\rangle}\left(\tilde{J}_{(1^4)}(\partial x)\tau\right)\cdot \tau\nn\\
&-&\frac{2(\alpha-1)}{2+\alpha}\frac{1}{\langle \tilde{J}_{(1^3)},\tilde{J}_{(1^3)}\rangle}\left(\tilde{J}_{(1^3)}(\partial x)\tau\right)\cdot \frac{1}{\langle \tilde{J}_{(1)},\tilde{J}_{(1)}\rangle}\left(\tilde{J}_{(1)}(\partial x)\tau\right)\nn\\
&-&\frac{2}{1+\alpha}\frac{1}{\langle \tilde{J}_{(2,1)},\tilde{J}_{(2,1)}\rangle}\left(\tilde{J}_{(2,1)}(\partial x)\tau\right)\cdot \frac{1}{\langle \tilde{J}_{(1)},\tilde{J}_{(1)}\rangle}\left(\tilde{J}_{(1)}(\partial x)\tau\right)\nn\\
&+&\frac{2}{(1+\alpha)}\frac{1}{\langle \tilde{J}_{(2)},\tilde{J}_{(2)}\rangle}\left(\tilde{J}_{(2)}(\partial x)\tau\right)\cdot \frac{1}{\langle \tilde{J}_{(1^2)},\tilde{J}_{(1^2)}\rangle}\left(\tilde{J}_{(1^2)}(\partial x)\tau\right)\nn\\
&+&\frac{(\alpha-1)}{ 1+\alpha}\frac{1}{\langle \tilde{J}_{(1^2)},\tilde{J}_{(1^2)}\rangle}\left(\tilde{J}_{(1^2)}(\partial x)\tau\right)\cdot \frac{1}{\langle \tilde{J}_{(1^2)},\tilde{J}_{(1^2)}\rangle}\left(\tilde{J}_{(1^2)}(\partial x)\tau\right)=0.\label{hbareqJack}
\eea
If we substitute the expressions of Jack polynomials $\tilde{J}_\lambda$, this equation becomes (\ref{hbarKP}) and (\ref{hbareqHirota}). The equation (\ref{hbareqJack}) can also be written as
\be
S_{(2,2)}(\hbar\partial x)\tau\cdot \tau-S_{(2,1)}(\hbar\partial x)\tau\cdot S_{(1)}(\hbar\partial x)\tau+S_{(2)}(\hbar\partial x)\tau \cdot S_{(1^2)}(\hbar\partial x)\tau=0.
\ee
We write the tau functions $\tau$ of the forms
\be
\tau=\sum_{\lambda}c_\lambda'\tilde{J}_\lambda(p)=\sum_{\lambda}c_\lambda S_\lambda\left(\frac{p}{\sqrt{\alpha}}\right).
\ee
If $c_\lambda$ or $c_\lambda'$ satisfy some relations (the Pl\"ucker relations), the tau functions $\tau$ are the solutions of the $\hbar$-KP hierarchy. For example,
\bea
c_{(2,2)}c_{\emptyset}-c_{(2,1)}c_{(1)}+c_{(2)}c_{(1,1)}=0.\label{clambdarelation}
\eea
In fact, the coefficients $c_\lambda$ satisfy the classical Pl\"ucker relations since $S_\lambda(\hbar \partial_p)S_\mu\left(\frac{p}{\hbar}\right)|_{p=0}=\delta_{\lambda\mu}$, which is the same with that in \cite{APSZ}. For $c_\lambda'$, since \[
\tilde{J}_\lambda(\partial_p)\tilde{J}_\mu(p)|_{p=0}=\delta_{\lambda\mu}\langle \tilde{J}_\lambda,\tilde{J}_\lambda\rangle,\]
we obtain the relations of $c_\lambda'$. For example,
\bea
&&\frac{4}{(1+\alpha)^2}c_{(2,2)}'c_{\emptyset}'+\frac{6(\alpha-1)}{(2+\alpha)(3+\alpha)}c_{(2,1,1)}'c_{\emptyset}'
+\frac{2(\alpha-1)\alpha}{(2+\alpha)(3+\alpha)}c_{(1^4)}'c_{\emptyset}'
-\frac{2(\alpha-1)}{2+\alpha}c_{(1^3)}'c_{(1)}'\nn\\
&&
-\frac{2}{1+\alpha}c_{(2,1)}'c_{(1)}'
+\frac{2}{(1+\alpha)}c_{(2)}'c_{(1^2)}'
+\frac{(\alpha-1)}{ 1+\alpha}c_{(1^2)}'c_{(1^2)}'=0.\label{clambda'relation}
\eea
The set of coefficients $\{c_\lambda\}$ can be represented linearly by the set $\{c_\lambda'\}$ and vice versa. In fact, $\{c_\lambda\}$ satisfies the classical Pl\"ucker relations if and only if $\{c_\lambda'\}$ satisfies its Pl\"ucker realtions. For example, $\{c_\lambda\}$ satisfies (\ref{clambdarelation}) if and only if $\{c_\lambda'\}$ satisfies (\ref{clambda'relation}).
\section{The Boson-Fermion correspondence for $\hbar$-KP hierarchy}\label{sect5}
Let $p_n=nx_n$. For Schur functions $S_\lambda(x)$ and Jack polynomials $\tilde{J}_{\lambda}(x)$, we have
\bea
e^{\sum_{n=1}^\infty x_n z^n}&=&\sum_{n\geq 0}S_{(n)}(x)z^n,\\
e^{-\sum_{n=1}^\infty x_n z^n}&=&\sum_{n\geq 0}(-1)^nS_{(1^n)}(x)z^n.
\eea
Then we have
\bea
\exp\left({\sum_{n=1}^\infty \frac{x_n}{\sqrt{\alpha}} z^n}\right)&=&\sum_{n\geq 0}S_{(n)}\left(\frac{x}{\sqrt{\alpha}}\right)z^n=\frac{1}{\langle \tilde{J}_{(n)},\tilde{J}_{(n)}\rangle\sqrt{\alpha}^n} \tilde{J}_{(n)}(x)z^n,\\
\exp\left(-\sum_{n\geq 1}x_n\sqrt{\alpha}z^{n}\right)&=&\sum_{n\geq 0}(-1)^nS_{1^n}(\sqrt{\alpha}x)z^n=\sum_{n=0}^\infty\frac{(-1)^n\sqrt{\alpha}^n}{\langle \tilde{J}_{(1^n)},\tilde{J}_{(1^n)}\rangle }\tilde{J}_{(1^n)}(p)z^n.
\eea
The Cauchy formula is
\bea
&&\exp\left(\sum_{n\geq 1}\frac{p_np_n'}{n}\right)=\exp(\sum_{n\geq 1} nx_nx_n')\nn\\
&=&\sum_{\lambda}S_{\lambda}(x)S_{\lambda}(x')=\sum_{\lambda}\frac{1}{\langle \tilde{J}_\lambda,\tilde{J}_\lambda\rangle}\tilde{J}_{\lambda}(x)\tilde{J}_{\lambda}(x').
\eea
Note that $\tilde{J}_\lambda(x)$ is not equal to $S_\lambda(x)$ (or multiplied by a constant).

The Fermions $\psi_j$ and $\psi_{j}^*$ are defined as usual. For $j\in\Z+\frac{1}{2}$, $\psi_j$ and $\psi_{j}^*$ satisfy\cite{MJD}
\be
[\psi_i,\psi_j]_+=0,\ [\psi_i^*,\psi_j^*]=0,\ [\psi_i,\psi_j^*]=\delta_{i+j,0},
\ee
where $[A,B]_+=AB-BA$. Particularly,
\[
\psi_j^2=0,\ \ \psi_j^{*2}=0.
\]
The Fermionic Fock space $\mathcal{F}$ is the space of Maya diagrams\cite{MJD}. A Maya diagram can be discribed as an increasing sequence of half-integers
\[
|{\bf u}\rangle=|u_1,u_2,\cdots\rangle,\ \ \text{with}\ \ u_1<u_2<\cdots,
\]
and $u_{j+1}=u_j+1$ for all sufficiently large $j$.

The actions of Fermions $\psi_j, \psi^*_j$ on Maya diagrams are determined by
\begin{eqnarray}
\psi_j|{\bf u}\rangle &=&  \begin{cases}
 (-1)^{i-1}| \cdots,u_{i-1},u_{i+1},\cdots\rangle & \text{ if } u_i=-j \ \text{for some} \ i, \\
  0& \text{ otherwise},
\end{cases}  \\
\psi^*_j |{\bf u}\rangle &=&  \begin{cases}
 (-1)^{i}| \cdots,u_{i},j,u_{i+1},\cdots\rangle & \text{ if } u_i<j<u_{i+1} \ \text{for some} \ i, \\
  0& \text{ otherwise}.
\end{cases}
\end{eqnarray}
The generating functions of Fermions are
\[
\psi(z)=\sum_{j\in\Z+1/2}\psi_jz^{-j-1/2},\ \psi^*(z)=\sum_{j\in\Z+1/2}\psi^*_jz^{-j-1/2}.
\]
The normal order is defined as usual. For Maya diagrams $|{\bf u}\rangle$ and $|{\bf v}\rangle$, the pair $\langle{\bf v}|{\bf u}\rangle$ is defined by the formula
\[
\langle{\bf v}|{\bf u}\rangle=\delta_{v_1+u_1,0}\delta_{v_2+u_2,0}\cdots.
\]
Let
\begin{equation}\label{hhx}
H_n=\sum_{j\in\Z+1/2}:\psi_{-j}\psi_{j+n}^*:.\end{equation}
It satisfy\cite{MJD}
\[
[H_n,\psi_j]=\psi_{n+j},\ [H_n,\psi_{j}^*]=-\psi_{n+j}^*,
\]
and
\[
[H_n, H_m]=n\delta_{n+m,0}.
\]
Then we show the Boson-Fermion correspondence in the $\hbar$-KP hierarchy.  Define
\be
H(x)=\sum_{n\geq 1}\frac{x_n}{\sqrt{\alpha}}H_n.
\ee
For any element $|{\bf{u}}\rangle\in \mathcal{F}$, define the map
\be
\Phi(|u\rangle)=\sum_{l\in\Z}z^l\langle l|e^{H(x)}|{\bf u}\rangle.
\ee
Then $\Phi(z)$ is in the space $\C(\alpha)[z,z^{-1},x_1,x_2,\cdots]$, and the correspondence $\Phi$ is an isomorphism of the vector spaces $\mathcal{F}$ over $\C(\alpha)$ and $\C(\alpha)[z,z^{-1},t_1,t_2,\cdots]$. Moreover, for $n>0$,
\be
\Phi(H_{n}|{\bf u}\rangle)=\sqrt{\alpha}\partial_{x_n}\Phi(|{\bf u}\rangle),\ \ \Phi(H_{-n}|{\bf u}\rangle)=n\frac{x_n}{\sqrt{\alpha}}\Phi(|{\bf u}\rangle).
\ee

From the commutation relations above, we get
\bea
[H(x),\psi(z)]&=&\left(\sum_{n\geq 1}\frac{x_n}{\sqrt{\alpha}}z^n\right)\psi(z),
\eea
and
\be
[H(x),\psi^*(z)]=\left(-\sum_{n\geq 1}\frac{x_n}{\sqrt{\alpha}}z^n\right)\psi^*(z),
\ee
which show
\bea
e^{H(x)}\psi(z)e^{-H(x)}=e^{\sum_{n\geq 1}\frac{x_n}{\sqrt{\alpha}}z^n}\psi(z),\\
e^{H(x)}\psi(z)^*e^{-H(x)}=e^{-\sum_{n\geq 1}\frac{x_n}{\sqrt{\alpha}}z^n}\psi^*(z).
\eea
Then
\bea
e^{H(x)}\psi_je^{-H(x)}&=&\sum_{n=1}^\infty\psi_{j+n}S_{(n)}\left(\frac{x}{\sqrt{\alpha}}\right),\\
e^{H(x)}\psi_j^*e^{-H(x)}&=&\sum_{n=1}^\infty\psi_{j+n}S_{(n)}\left(-\frac{x}{\sqrt{\alpha}}\right).
\eea
From these formulas, we polynomials $\Phi(|{\bf u}\rangle)$ can be determined. For example, let $|{\bf u}\rangle=\psi_{-5/2}|\text{vac}\rangle$,
\beaa
\Phi(\psi_{-5/2}|\text{vac}\rangle)&=&z\langle 1|e^{H(x)}\psi_{-5/2}|\text{vac}\rangle\\
&=&z\langle \text{vac}|\psi_{1/2}^*e^{H(x)}\psi_{-5/2}e^{-H(x)}|\text{vac}\rangle\\
&=&z S_{(2)}\left(\frac{x}{\sqrt{\alpha}}\right)\\
&=&z\frac{1}{\langle \tilde{J}_{(2)},\tilde{J}_{(2)}\rangle \sqrt{\alpha}^2}\tilde{J}_{(2)}(x).
\eeaa

Let the operators $z^{H_0}$ and $e^K$ are defined the same as that in \cite{MJD}. Then define
\bea
\Psi(z)&=&\sum_{j\in\Z+\frac{1}{2}}\Psi_jz^{-j-1/2}=e^{\left(\sum_{n\geq 1}\frac{x_n}{\sqrt{\alpha}}z^{n}\right)}e^{\left(-\sum_{n\geq 1}{\partial_{p_n}}\sqrt{\alpha}z^{-n}\right)}e^Kz^{H_0}
\label{vertexv+zxn},\\
\Psi^*(z)&=&\sum_{j\in\Z+\frac{1}{2}}\Psi_j^*z^{-j-1/2}=e^{\left(-\sum_{n\geq 1}\frac{x_n}{\sqrt{\alpha}}z^{n}\right)}e^{\left(\sum_{n\geq 1}{\partial_{p_n}}\sqrt{\alpha}z^{-n}\right)}e^{-K}z^{-H_0}.\label{vertexv-zpn}
\eea
They give the realization of the Fermioinic generating functions $\psi(z)$ and $\psi^*(z)$ in the Bosonic Fock space $\C(\alpha)[z,z^{-1},t_1,t_2,\cdots]$, that is, we have
\bea
\Phi(\psi(z)|{\bf u}\rangle)=\Psi(z)\Phi(|{\bf{u}}\rangle),\ \Phi(\psi^*(z)|{\bf u}\rangle)=\Psi^*(z)\Phi(|{\bf{u}}\rangle).
\eea
For example,
\beaa
\Psi_{-5/2}\cdot 1&=&\sum_{n-m=2} S_{(n)}\left(\frac{x}{\sqrt{\alpha}}\right)(-1)^m S_{(1^m)}(\sqrt{\alpha}\partial_x)\cdot 1=S_{(2)}\left(\frac{x}{\sqrt{\alpha}}\right),
\eeaa
which equals $\Phi(\psi_{-5/2}|\text{vac}\rangle)$.
Then the $\hbar$-KP hierarchy can be written in the Fermion form
\be
\sum_{j\in\Z+\frac{1}{2}}\psi_j^*\tau\otimes \psi_{-j}\tau=0.
\ee

 \section*{Data availability statement}
The data that support the findings of this study are available from the corresponding author upon reasonable request.

\section*{Declaration of interest statement}
The authors declare that we have no known competing financial interests or personal relationships that could have appeared to influence the work reported in this paper.

\section*{Acknowledgements}
This research is supported by the National Natural Science Foundation
of China under Grant No. 12101184 and No. 11871350, and supported by the Key Scientific Research Project in Colleges and Universities of Henan Province No. 22B110003.


\begin{thebibliography}{100}
\bibitem{DKJM}E. Date, M. Kashiwara, M. Jimbo, T. Miwa, {\it Transformation groups for soliton equations.} Nonlinear integrable systems-classical theory and quantum theory (Kyoto,1981), World Sci. Publishing, Singapore, 1983, 39-119.
\bibitem{MJD}
 T. Miwa, M. Jimbo, E. Date, {\it Solitons: Differential equations, symmetries and infinite dimensional algebras}. Cambridge University Press, Cambridge, 2000.
 \bibitem{FH} W. Fulton, J. Harris, Representation theory, A first course. Springer-Verlag, New York, 1991.
\bibitem{Mac} I. G. Macdonald, Symmetric functions and Hall polynomials. Oxford Mathematical Monographs, Clarendon Press, Oxford, 1979.
\bibitem{weyl}H. Weyl, \emph{The classical groups; their invariants and representations}. Princeton Univ. Press, Princeton, 1946.
\bibitem{Stan} R. P. Stanley, Enumerative Combinatorics, Volume II. Cambridge University Press, Cambridge, 1999.

\bibitem{TT1} K. Takasaki, T. Takebe, \emph{Integrable hierarchies and dispersionless limit.} Rev. Math. Phys.,
07 (1995), 743-808.

\bibitem{TT2} K. Takasaki, T. Takebe, \emph{$\hbar$-dependent KP hierarchy.} Theor. Math. Phys.,
171 (2012), 683-690.

\bibitem{deformedKPJack}
N. Wang, \emph{Jack polynomials, the deformed KP hirarchy and affine Yangian of ${\mathfrak{gl}}(1)$}. submitted.

\bibitem{Pro} T. Proch\'{a}zka, $\mathcal{W}$-symmetry, topological vertex and affine Yangian, JHEP 10 (2016) 077.
\bibitem{Tsy} A. Tsymbaliuk, The affine Yangian of $gl_1$ revisited, Adv. Math. 304 (2017) 583-645, arXiv:1404.5240.

\bibitem{FW} O. Foda, M. Wheeler. {\it Hall-Littlewood plane partitions and KP},	Int. Math. Res. Not \textbf{26}(2), (2009) 2597.
\bibitem{ORV}
A. Okounkov, N. Reshetikhin, C. Vafa. {\it Quantum Calabi-Yau and classical crystals},  \url{arXiv:hep-th/0309208}.
\bibitem{NT}
T. Nakatsu, K. Takasaki. {\it Integrable structure of melting crystal model with external potentials}, Adv. Stud. Pure  Math \textbf{26}(2), 59 (2010) 201-223.
\bibitem{3DFermionYangian}
N. Wang, K. Wu, \emph{3D Fermion Representation of Affine Yangian}, Nucl. phys. B 974 (2022) 115642.
\bibitem{JackYangian} Z. Cui, Y. Bai, N. Wang, K. Wu, {\it Jack polynomials and affine Yangian}, Nucl. phys. B 984 (2022) 115986.
\bibitem{WBCW} N. Wang, B. Yang, Z. N. Cui, K. Wu, \emph{Symmetric functions and 3D Fermion representation of $W_{1+\infty}$ algebra}, accepted to Advances in Applied Clifford Algebras.

\bibitem{APSZ}A. Andreev, A. Popolitov, A. Sleptsov, A. Zhabin, \emph{Genus expansion of matrix models and $\hbar$ expansion of KP hierarchy}. JHEP, 12 (2020) 038.




\end{thebibliography}
\end{document}